# Human-AI Interaction Design Standards

Chaoyi Zhao [0000-0002-7386-6847]

Laboratory of Human Factors and Ergonomics, China National Institute of Standardization

4 Zhichun Road, Haidian District, Beijing 100191, China,

e-mail: zhaochy@cnis.ac.cn

Wei Xu

Department of Psychology and Behavioral Sciences, the Center for Psychological Sciences, Zhejiang University

866 Yuhangtang Rd, Xihu, Hangzhou, Zhejiang, China, 310027

Email: weixu6@yahoo.com

## Abstract

The rapid development of artificial intelligence (AI) has significantly transformed human-computer interactions, making it essential to establish robust design standards to ensure effective, ethical, and human-centered AI (HCAI) solutions. Standards serve as the foundation for the widespread adoption of new technologies, and human-AI interaction (HAII) standards are critical to supporting the industrialization of AI technology by following an HCAI approach. These design standards aim to provide clear principles, requirements, and guidelines for designing, developing, deploying, and using AI systems, enhancing the user experience and performance of AI systems. Despite their importance, the creation and adoption of HCAI-based interaction design standards face several challenges, including the absence of universal frameworks, the inherent complexity of HAII, and the ethical dilemmas that arise in such systems. This chapter provides a comparative analysis of Human-AI Interaction (HAII) versus traditional human-computer interaction (HCI) and outlines guiding principles for HCAI-based design. It explores international, regional, national, and industry standards related to HAII design from an HCAI perspective and reviews design guidelines released by leading companies such as Microsoft, Google, and Apple. Additionally, the chapter highlights tools available for implementing HAII standards and presents case studies of human-centered interaction design for AI systems in diverse fields, including healthcare, autonomous vehicles, and customer service. It further examines key challenges in developing HAII standards and suggests future directions for the field. Emphasizing the importance of ongoing collaboration between AI designers, developers, and experts in human factors and human-computer interaction (HCI), this chapter stresses the need to advance HCAI-based interaction design standards to ensure human-centered AI solutions across various domains.

## Keywords

# 1 Introduction

Human-centered artificial intelligence (HCAI) is an approach to artificial intelligence design and development that emphasizes the augmentation of human capabilities, user empowerment, and ethical considerations. It seeks to align artificial intelligence (AI) systems with human needs, values, and expectations to ensure that technology serves individuals and society. HCAI systems are designed to prioritize usability, transparency, and collaboration, enabling humans to retain control and trust over AI-powered solutions (Xu, 2019; Shneiderman, 2020). The key principles of HCAI include enhancing human decision-making, adhering to ethical and fairness standards, and ensuring transparency through explainable AI systems.

The interaction between humans and AI systems plays a crucial role in determining the effectiveness, acceptance, and societal impact of AI technologies. Properly designed human-AI interactions (HAII) improve user experience and accessibility by bridging the gap between AI systems and end-users. By providing intuitive user interfaces, multimodal interactions (e.g., voice, gesture, or touch), and clear feedback mechanisms, users are more likely to engage with and benefit from AI systems (Adam & Wessel, 2020). Furthermore, trust is fundamental in AI adoption, and effective interaction design fosters this trust by integrating explainable AI (XAI) features. These features help users understand how decisions are made, thereby increasing confidence in the system (Doshi-Velez & Kim, 2017). Lastly, interaction design enables effective human-AI collaboration, where AI systems complement human expertise, enhancing decision-making in areas like healthcare, education, and customer service.

Despite its potential, HAII faces significant challenges. A primary issue is the lack of transparency and explainability in AI systems, particularly those using deep learning models. Often perceived as "black boxes," these systems make it difficult for users to understand their decision-making processes, leading to mistrust and reluctance in adoption (Shneiderman, 2020). Ethical and fairness concerns also persist, as AI systems may unintentionally perpetuate biases in their training data. For example, biased training sets can result in unfair or discriminatory outcomes, particularly in high-stakes applications like hiring or criminal justice (Adam & Wessel, 2020). Another challenge involves balancing automation with human control. Excessive reliance on automation can result in automation complacency, where users blindly trust the system, while systems requiring too much manual intervention can reduce efficiency (Goodall, 2014). Cognitive load and usability issues are also prevalent, as poorly designed user interfaces and excessive information can overwhelm users. Simplifying interactions without sacrificing functionality is a persistent design challenge (Norman, 2013). Furthermore, privacy and data security concerns arise due to the vast amount of personal data AI systems require for functionality. Users are increasingly concerned about how their data is collected, stored, and utilized, necessitating transparent data policies (Doshi-Velez & Kim, 2017). Societal and cultural differences present another layer of complexity. AI systems must adapt to diverse cultural norms and values to avoid misinterpretation or cultural insensitivity, particularly in global applications. Finally, seamless integration across devices and ecosystems is critical but difficult to achieve. Fragmentation in AI ecosystems can lead to poor user experiences and inefficiencies (Shneiderman, 2020).

Standards are fundamental to the widespread adoption of new technologies, and HAII standards play a crucial role in supporting the industrialization of AI technology. In response to the unique characteristics of HAII, various standards organizations have been actively developing HAII design standards (ISO, 2020; Schlenoff, 2022). Additionally, major technology companies such as Microsoft, Google, and Apple have published their own guidelines for HAII design (Amershi et al., 2019; Wright et al., 2020; Google, 2022; Apple, 2022). This chapter provides an overview of the foundations of standardization, including the different levels of standards. It summarizes the international HAII design standards released by the International Organization for Standardization (ISO) and the International Electrotechnical Commission (IEC), along with industry standards developed by the Institute of Electrical and Electronics Engineers (IEEE). It also reviews related national standards from countries such as the United Kingdom, the United States, and China. Furthermore, it outlines the HAII design guidelines introduced by Microsoft, Google, and Apple and offers a comparative analysis of HAII versus human-non-AI computer interaction. The chapter also explores tools available for implementing HAII standards and provides case studies of human-centered AI interaction design from various fields, including healthcare, autonomous vehicles, and customer service. Finally, it examines the key challenges in developing HCAI-based design standards and proposes future directions for advancing the field. By emphasizing the importance of ongoing collaboration between AI designers, developers, and human factors and HCI experts, the chapter highlights the necessity of creating HCAI-based interaction design standards to ensure ethical, effective, and user-friendly integration of AI systems across industries.

# 2 Comparative Analysis of HAII and Human-Non-AI Computer Interaction

Human-Computer Interaction (HCI) is a long-established field focusing on the design, evaluation, and implementation of user interfaces for non-AI computational systems. These systems, while sophisticated, typically lack the autonomy and adaptability characteristic of AI. The emergence of AI-powered systems has led to a new paradigm: HAII, where intelligent systems collaborate with users in dynamic, context-aware ways. Understanding the distinctions between these paradigms is critical for advancing user experience, human factors, and system design.

The shift from HCI to HAII represents a fundamental transformation in how humans interact with technology (Xu, W., Dainoff, M., Ge, L., Gao, Z., 2022). While HAII provides significant benefits such as automation and adaptability, it also introduces new challenges, including ethical considerations, trust management, and system transparency. Non-AI systems remain relevant in contexts requiring high user control and predictability, while AI-driven systems excel in dynamic and complex environments.

## 2.1 Adaptability and Learning

Adaptability and learning are the most striking differences between HAII and Human-Non-AI Computer Interaction (HCI). Traditional non-AI systems follow deterministic, rule-based programming, relying entirely on human developers to define their behavior. This inflexibility means non-AI systems perform well in structured tasks with predictable inputs but struggle when applied to dynamic or ambiguous scenarios. For instance, a non-AI word processor like Microsoft Word can only execute predefined operations, such as spell-checking or formatting, based on static algorithms. If a user desires a novel feature or encounters a complex scenario, the system cannot "learn" or adapt; instead, developers must update the software to accommodate such needs. In contrast, AI systems are built on machine learning (ML) frameworks, enabling them to analyze data, identify patterns, and adapt over time. For example, AI-powered applications like Grammarly not only detect spelling and grammatical errors but also learn user-specific preferences, such as tone, style, or vocabulary choices. This adaptability makes AI systems more suited for evolving tasks and environments. However, the reliance on ML also introduces challenges. AI systems often require vast datasets to achieve high adaptability, raising concerns about data privacy and algorithmic bias. Moreover, the learning process is probabilistic rather than deterministic, meaning the system might make errors during adaptation. In non-AI systems, errors are often predictable and traceable, but in AI systems, they may emerge unpredictably as the model processes novel data. Thus, while adaptability is a hallmark advantage of HAII, it also demands careful oversight to ensure reliability and fairness. (Norman, 2013; Amershi et al., 2019).

## 2.2 Decision-Making and Autonomy

Decision-making is another critical point of divergence between HAII and HCI. In traditional HCI systems, decision-making lies exclusively with the user. The system acts as a passive tool, providing functionalities and waiting for user commands. For example, in a spreadsheet application such as Microsoft Excel, the user defines every operation, whether it's calculating sums, creating graphs, or applying conditional formatting. The system merely executes these commands without independent analysis or inference. This explicit control ensures that the user remains the primary decision-maker, which is often desirable in environments where predictability and accountability are paramount. On the other hand, AI-powered systems introduce a new dimension of autonomy. AI systems can independently process large datasets, identify trends, and make recommendations or decisions without explicit user instructions. For example, financial AI systems can analyze market trends in real time, predict stock performance, and autonomously execute trades. While this autonomy reduces user workload and enhances efficiency, it also introduces risks. A lack of transparency in AI decision-making processes—often referred to as the "black-box" problem—can lead to distrust or ethical concerns. Users may struggle to understand or control the AI's decisions, particularly if these decisions conflict with user expectations or values. Additionally, errors in AI decision-making may result from biased training data or flawed algorithms, further complicating accountability. As AI continues to gain autonomy, striking a balance between user oversight and system independence remains a key challenge for designers and policymakers. (Amershi et al., 2019; Binns, 2018). The HCAI approach ensures that humans are the ultimate authority in decision-making (Xu, 2019).

## 2.3 User Experience and Interaction Style

User experience (UX) in Human-Non-AI Computer Interaction is traditionally structured and task-oriented. User interfaces are designed for consistency and clarity, focusing on usability principles such as ease of learning, efficiency, and error prevention. For example, enterprise software often uses menu-driven interfaces and predictable workflows, allowing users to accomplish tasks without needing extensive training. Interaction styles in such systems typically involve direct manipulation of controls, such as clicking buttons, entering commands, or navigating structured menus. These interactions are rigid and lack contextual awareness, requiring users to adapt to the system's constraints. In contrast, HAII introduces fluid and dynamic interaction styles. Many AI systems employ natural language processing (NLP) and multimodal interfaces, enabling users to communicate with systems conversationally. For instance, AI-powered virtual assistants like Siri or Alexa allow users to issue voice commands or ask questions in natural language. This conversational interface creates a more intuitive experience, lowering the cognitive load on users. Moreover, AI systems can personalize interactions by learning user preferences and intentions and adapting their responses accordingly. However, this flexibility comes with challenges. Over-reliance on adaptive AI systems can lead to inconsistent UX, as the system's behavior may vary unpredictably based on contextual factors. Additionally, users may struggle to understand the AI's reasoning, leading to potential frustration or mistrust. Designing AI-based user interfaces requires balancing adaptability with consistency, ensuring users feel empowered rather than overwhelmed by the system's capabilities. (Norman, 2013; Amershi et al., 2019).

## 2.4 Error Handling and Feedback

Error handling and feedback mechanisms significantly differ between HAII-based and HCI-based systems. In traditional HCI, error handling is deterministic, with systems following pre-defined rules to address errors. Feedback is provided through explicit messages, such as error codes or system prompts, which guide users to resolve issues. For example, when a user enters invalid input into a web form, the system highlights the error and provides a specific message like "Invalid email format." These interactions are straightforward, ensuring users can easily identify and correct mistakes. On the other hand, HAII systems introduce a probabilistic element to error handling. AI systems often generate predictions or decisions based on incomplete or noisy data, leading to errors that are less predictable. For instance, an AI-powered facial recognition system might misidentify individuals due to biased training data or poor lighting conditions. Feedback mechanisms in AI systems are more iterative, relying on user input or additional data to refine their models. This iterative process allows AI systems to "learn" from their mistakes, improving performance over time. However, the probabilistic nature of AI errors can frustrate users, particularly when errors are opaque or difficult to diagnose. Furthermore, biased or incomplete training data can lead to systemic errors that persist despite iterative learning. Effective error handling in HAII requires transparency, ensuring users understand the system's limitations and have mechanisms to correct or override erroneous decisions. Designing robust feedback loops that balance automation with user control is essential for building trust and usability in AI systems. (Amershi et al., 2019; Binns, 2018).

## 2.5 Ethical and Trust Implications

The ethical and trust implications of HAII far surpass those of traditional HCI systems. In HCI, ethical concerns typically revolve around data security, accessibility, and usability. Trust is built on the predictability and transparency of system behavior. For example, a non-AI email client like Microsoft Outlook adheres to strict user-defined rules for managing emails, ensuring predictable outcomes. However, in HAII, ethical considerations become more complex. AI systems often process vast amounts of personal data to deliver personalized experiences, raising privacy concerns. Moreover, biases in training data can lead to unfair or discriminatory outcomes, particularly in sensitive applications like hiring, healthcare, or criminal justice. For instance, an AI-powered hiring platform might inadvertently favor candidates from certain demographic groups due to biased training data, perpetuating inequality. Trust in AI systems depends on their transparency, explainability, and fairness. Users need to understand how decisions are made and feel confident that the AI aligns with their values and goals. However, the "black-box" nature of many AI systems complicates this process. Efforts to build explainable AI (XAI) aim to address these challenges by making AI decision-making processes more interpretable and accountable. Designers and developers must prioritize ethical considerations, ensuring AI systems respect user autonomy, minimize bias, and promote fairness. Addressing these challenges is crucial for fostering trust and ensuring the responsible deployment of AI technologies. (Binns, 2018; Amershi et al., 2019).

Given the emerging characteristics of AI technology, as discussed earlier, the development of HAII design standards necessitates innovative design thinking and approaches. At the same time, these standards are essential to addressing the unique challenges posed by AI technologies, ensuring the creation of truly HCAI systems.

# 3 HCAI-based design guiding principles

Human-Centered AI (HCAI) interaction focuses on designing systems that prioritize user needs, ethical standards, and usability while fostering trust and collaboration between humans and AI. To achieve these goals, effective HAII design is essential. Such designs must ensure usability, build trust, and align with ethical principles, enabling AI systems to complement human capabilities. The key design guiding principles for effective HAII, with a focus on transparency, usability, personalization, and accountability, are outlined below.

## 3.1 Transparency and Explainability

Transparency and explainability are fundamental to fostering trust and confidence in AI systems. Users need to understand how AI systems make decisions, especially in high-stakes scenarios such as healthcare, finance, or autonomous driving. Explainable AI (XAI) features play a critical role in demystifying complex algorithms by providing clear and concise explanations of system processes, inputs, and outputs. For instance, visual interfaces can present decision pathways or contributing factors in user-friendly formats, enabling non-technical users to grasp how the system operates. Additionally, systems must communicate their limitations, such as when predictions or recommendations are uncertain or based on incomplete data. This builds realistic expectations and reduces the risk of over-reliance. Transparency also involves proactive communication about ethical considerations, including how user data is collected and processed. By addressing these aspects, designers can ensure that users feel informed, empowered, and confident in their interactions with AI. Ultimately, transparent and explainable systems not only enhance user trust but also promote accountability, as they allow users to audit and challenge decisions when necessary, fostering a collaborative relationship between humans and AI.

## 3.2 Usability and Accessibility

Usability and accessibility are critical for ensuring that AI systems can be effectively and comfortably used by a diverse range of users. A user-friendly system minimizes the cognitive load required to operate it and ensures that users can accomplish their goals with minimal effort. Intuitive design principles, such as clear navigation, logical grouping of features, and consistent visual cues, are essential to improving usability. Accessibility, on the other hand, ensures that AI systems can cater to people with varying levels of ability, including those with disabilities. For example, multimodal interfaces—such as voice commands for visually impaired users or haptic feedback for those with hearing impairments—broaden access to AI technology. Comprehensive user testing with individuals from diverse backgrounds is crucial to identify potential usability barriers and address them before deployment. Additionally, systems should provide onboarding tutorials, context-sensitive help features, and troubleshooting guides to assist users in navigating the system effectively. By prioritizing usability and accessibility, AI systems can provide inclusive and seamless experiences, ensuring that users of all backgrounds can benefit equally from technological advancements.

## 3.3 Personalization and Adaptability

Personalization and adaptability enhance the relevance and value of AI systems by tailoring their functionality to the unique needs, preferences, and behaviors of individual users. Personalization allows users to customize system settings and interaction styles, such as choosing preferred notification methods, adjusting interface themes, or configuring automated routines. Machine learning algorithms can further improve adaptability by analyzing user behavior and preferences over time, enabling the system to predict and fulfil user needs proactively. For example, a smart thermostat can learn a household's temperature preferences and automatically adjust settings based on time of day or weather conditions. Supporting multiple user profiles is another crucial aspect of personalization, particularly in shared environments, where each user may have distinct requirements. However, personalization must strike a balance between user convenience and privacy, ensuring that sensitive data is handled transparently and securely. By delivering tailored experiences that evolve with user needs, personalized AI systems foster engagement, satisfaction, and long-term adoption while reinforcing the perception that the technology truly serves the individual.

## 3.4 Ethical Alignment and Fairness

Ethical alignment and fairness are central to designing responsible AI systems that respect societal norms and avoid perpetuating harm. AI systems must be developed and deployed in ways that prevent bias, discrimination, or unethical outcomes, particularly in sensitive applications like hiring, criminal justice, or healthcare. Regular audits and evaluations of AI models are essential to identifying and addressing potential biases in datasets, algorithms, and decision-making processes. Designers must ensure that datasets used to train AI models are diverse, representative, and free from historical biases that could influence outcomes unfairly. Moreover, systems should provide users with tools to review, challenge, or override decisions that may appear biased or inaccurate. Transparent communication about the ethical principles guiding the system, such as prioritizing equity and inclusivity, is also necessary to build user trust. By prioritizing fairness and ethical considerations, designers can ensure that AI systems align with societal values, promote equal opportunities, and avoid unintended negative consequences, ultimately contributing to a more equitable and just technology landscape.

## 3.5 Trust and Reliability

Trust and reliability are essential for encouraging user adoption and ensuring that AI systems function as dependable tools in various domains. Reliable systems provide consistent and accurate results, avoiding unexpected errors that could undermine user confidence. Trust can be built by ensuring the system behaves predictably and by offering users clear, timely feedback on its operations. For instance, users should receive updates when the system successfully completes a task or when an error occurs, accompanied by explanations of why the issue happened and how it can be resolved. Notifications about software updates, changes in system performance, or potential risks also contribute to trust. In addition, incorporating mechanisms that allow users to verify system outputs or test its functionality reinforces confidence in the AI's capabilities. Trust is further enhanced when AI systems are transparent about their limitations and designed with safeguards to prevent over-reliance, such as prompting users to confirm critical actions. By focusing on trust and reliability, designers can create systems that users feel comfortable relying on, even in high-stakes or critical scenarios.

## 3.6 Collaboration and Control

Effective collaboration and control are vital to ensuring that users and AI systems work together harmoniously. AI systems should be designed to augment human capabilities, not replace them, by facilitating a balanced partnership. This involves giving users clear and intuitive options to take control when necessary while allowing the AI to handle repetitive or time-consuming tasks. For instance, in applications like autonomous vehicles, AI systems must seamlessly transition control between the human driver and the automation based on context, such as during emergencies or complex driving conditions. Collaboration also requires clear communication, with the system providing feedback on its actions and decisions in a way that keeps users informed and engaged. For example, an AI assistant scheduling meetings should alert the user to potential conflicts and suggest alternative solutions, empowering the user to make the final decision. By designing systems that respect user autonomy and encourage participation, HAII can become more effective, with users feeling confident in their ability to guide and collaborate with AI.

## 3.7 Privacy and Data Security

Privacy and data security are critical aspects of HAII design, as users increasingly rely on AI systems that handle sensitive personal information. To maintain user trust, AI systems must adhere to data protection regulations, such as GDPR or CCPA, and implement robust security measures like encryption and secure authentication protocols. Transparency about data usage is equally important; users should be informed about what data is being collected, how it is stored, and how it will be used. For example, providing users with control over data sharing settings—such as the ability to opt out of certain data collection practices—empowers them to make informed decisions about their privacy. Designers should also implement anonymization techniques to protect user identities and ensure that sensitive information is not exposed during data processing. Regular audits and security updates are necessary to address emerging threats and maintain system integrity. By prioritizing privacy and security, AI systems can foster a sense of safety and trust, enabling users to interact with confidence and peace of mind.

# 4 Standards System Overview

## 4.1 Concept and Role of Standards

A standard is a document, established by consensus and approved by a recognized body, that provides, for common and repeated use, rules, guidelines, or characteristics for activities or their results, aimed at the achievement of the optimum degree of order in a given context (ISO/IEC, 2004). Standards should be based on the consolidated results of science, technology and experience, and aimed at the promotion of optimum community benefits.

Standardization is the process of obtaining the best order and benefits for repetitive things and concepts through the development and implementation of standards. As the main technical basis for economic and social activities, standardization promotes the best practices in industry sectors, can improve the applicability of products, processes, and services; is conducive to accelerating technological innovation and results transformation; helps to enhance the quality of products and services; helps to ensure user safety. The adoption of international standards can effectively prevent trade barriers. Standardization work plays an important leading role in promoting technological innovation and supporting industrial development.

## 4.2 Levels of Standards

According to the scope of standardization activities, the levels of standards can be divided into international standards, regional standards, national standards, industry standards, and corporate standards, which are issued by different levels of organizational institutions.

International standards provide a common global framework for practices, ensuring consistency and compatibility across countries, and serve as the highest level of standards, guiding industries worldwide. Regional standards may be harmonized with international standards but are often tailored to address specific regional issues. National standards may align with international and regional standards but are adjusted to the country's legal, economic, and cultural context. While industry standards may align with national or international standards, they often contain more detailed technical specifications tailored to the industry's practices, technology, or innovations. Corporate standards are typically based on or reference international, regional, national, and industry standards, are mandatory within the company and help ensure consistency and quality across its operations. They may be more detailed and specific than external standards, reflecting the company's unique requirements, goals, or competitive advantages. These various levels of standards are interconnected and designed to complement each other, providing a cohesive regulatory framework that balances global consistency with local relevance and sector-specific needs.

## 4.2.1 International Standards

International standards are developed by international organizations such as the International Organization for Standardization (ISO), the International Electrotechnical Commission (IEC), or the International Telecommunication Union (ITU). They ensure global compatibility, interoperability, and quality. International standards are essential for products or services that are traded across borders. They serve as the highest level of standards, guiding industries worldwide. International standards often serve as the foundation for regional, national, and industry standards, which in turn influence corporate standards.

In addition to ISO, IEC, ITU, organizations or institutions that issue international standards within a certain professional scope include the International Bureau of Weights and Measures (BIPM), Codex Alimentarius Commission (CAC), International Atomic Energy Agency (IAEA), International Maritime Organization (IMO), World Health Organization (WHO), etc.

## 4.2.2 Regional Standards

Regional standards are developed by organizations specific to a particular region. These standards take into account the specific needs, economic conditions, and regulatory environments of the region. Regional standards may be harmonized with international standards but are often tailored to address specific regional issues. They can help facilitate trade within the region and simplify compliance with local regulations. The main regional standards include the following.

European region: The main ones are the European Committee for Standardization (CEN), the European Committee for Electrotechnical Standardization (CENELEC), and the European Telecommunications Standards Institute (ETSI), whose mission is to develop and implement a coherent and consistent set of voluntary standards (Wetting, 2002) as the basis for the single European market/European Economic Area. For example, European Standards (EN) developed by CEN and CENELEC; European Standards in the field of telecommunications (ETSI EN) and European Telecommunications Standards Institute Standards (ETSI ES) developed by ETSI.

Americas region: For example, Pan American Standards (COPANT) developed by the Pan American Standards Commission (COPANT).

Asia, Africa, and Arab regions: For example, standards (SARC) developed by the South Asian Regional Standards Organization (SARSO); standards (ARS) developed by the African Regional Standards Organization (ARSO); standards (ASMO) developed by the Arab Standards and Metrology Organization (ASMO).

### 4.2.3  National Standards

National standards are developed by standardization bodies within a specific country. These standards are mandatory or voluntary rules governing local industries. National standards may align with international and regional standards but are adjusted to the country's legal, economic, and cultural context. They ensure local product safety, environmental protection, and consumer rights.

Many developed countries have their own national standardization body, the main national standards are: American National Standards (ANS) developed by the American National Standards Institute (ANSI), British Standards (BS) developed by the British Standards Institution (BSI), German Standards (DIN) developed by the German Institute for Standardization (DIN), French Standards (NF) developed by the French Standardization Association (AFNOR), Canadian Standards (CAN) developed by the Standards Council of Canada (SCC), Japanese Industrial Standards (JIS) developed by the Japanese Industrial Standards Committee (JISC), and Chinese National Standards (GB/T) developed by the Standardization Administration of China (SAC).

### 4.2.4  Industry standards

Industry standards are developed by specific sectors or industries, often by trade associations or professional bodies. These standards are more specialized and address the needs of particular industries (e.g., automotive, telecommunications, or food safety). While industry standards may align with national or international standards, they often contain more detailed technical specifications tailored to the industry's practices, technology, or innovations.

Some associations have developed some widely influential standards in certain professional fields, such as the American Society for Testing and Materials (ASTM), the American Society of Mechanical Engineers (ASME) , IEEE, the Society of Automotive Engineers (SAE), the World Wide Web Consortium (W3C), etc.

### 4.2.5  Corporate standards

Corporate standards are created within a specific company or organization to meet its operational needs and ensure product or service quality. These standards are typically based on or reference international, regional, national, and industry standards. Corporate standards are mandatory within the company and help ensure consistency and quality across its operations. They may be more detailed and specific than external standards, reflecting the company's unique requirements, goals, or competitive advantages. Some corporate standards are also widely used and become de facto industry standards, such as corporate standards established by Underwriters Laboratories (UL), Microsoft Corporation, Google Inc., etc.

## 4.3  Common Process for Developing Standards

The process of developing standards typically follows a structured approach to ensure consensus, quality, and broad applicability. Below are the key steps in the common process.

### 4.3.1  Identification of Need

A need for a standard is identified by stakeholders such as industry groups, governments, or organizations. This may arise from issues like safety concerns, market demand, or technological advancements.

### 4.3.2 Proposal and Approval

A formal proposal for the standard is submitted to a recognized standardization body (e.g., ISO, IEC, IEEE, or national bodies like ANSI). The proposal is reviewed and approved to move forward.

### 4.3.3 Formation of a Technical Committee

A technical committee (TC) or working group (WG) comprising experts, stakeholders, and representatives from various sectors is formed to develop the standard.

### 4.3.4 Drafting of the Standard

The committee drafts the standard, including technical specifications, guidelines, and requirements. This process involves collaboration, research, and consultation.

### 4.3.5 Public Review and Consultation

The draft standard is made available for public review, allowing stakeholders and the general public to provide feedback. This step ensures transparency and inclusivity.

### 4.3.6 Revision and Finalization

Based on feedback, the draft is revised and finalized. The committee ensures that the standard reflects consensus among stakeholders.

### 4.3.7 Approval and Publication

The finalized standard is submitted for formal approval by the standardization body. Once approved, it is published and made publicly available.

### 4.3.8 Implementation and Maintenance

Organizations and industries implement the standard. Periodic reviews and updates ensure the standard remains relevant and aligned with new developments or needs.

# 5 Overview of Standards for HAII Design

## 5.1 International Standards

### 5.1.1 ISO/TC159/SC 4 Ergonomics of human-system interaction subcommittee

The Ergonomics of Human-System Interaction Subcommittee (ISO/TC 159/SC 4), under the ISO/TC 159 Ergonomics Technical Committee, is responsible for developing standards related to the ergonomics of human-system interaction. These systems are typically computer-based and include hardware ergonomics (e.g., input devices, display, and interaction devices), software ergonomics (e.g., interaction and interface design), and use environment ergonomics (e.g., tasks, environment, and workplace). Additionally, the subcommittee focuses on human-centered design processes and methods. The general principles of ergonomic design for interactive systems and information presentation, as developed by ISO/TC 159/SC 4, are independent of specific usage environments or technical conditions and are applicable to all types of interactive systems. These principles and requirements are typically not tied to specific design styles or application domains. The main international standards applicable to HAII design are introduced briefly as follows.

ISO 9241-110: 2020 "*Ergonomics of human-system interaction — Part 110: Interaction principles*" describes principles for interaction between a user and a system that are formulated in general terms (i.e. independent of situations of use, application, environment or technology). This standard provides a framework for applying those interaction principles and the general design recommendations for interactive systems.

ISO 9241-112: 2017 "*Ergonomics of human-system interaction - Part 112: Principles of information presentation*" establishes ergonomic design principles for interactive systems related to the software-controlled presentation of information by user interfaces. It applies to the three main modalities (visual, auditory, tactile/haptic) typically used in information and communication technology. These principles apply to the perception and understanding of presented information. These principles are applicable in the analysis, design, and evaluation of interactive systems. This document also provides recommendations corresponding to the principles.

ISO 9241-115: 2024 "*Ergonomics of human-system interaction - Part 115: Guidance on conceptual design, user-system interaction design, user interface design and navigation design*" provides guidance on aspects of the design of human-system interaction, including conceptual design, user-system interaction design, user interface design and navigation design for interactive systems. This document applies to all design and development approaches and methodologies, including human-centred design, object-oriented, waterfall, human factors integration (HFI), agile and rapid development.

ISO 9241-129: 2010 "*Ergonomics of human-system interaction - Part 129: Guidance on software individualization*" provides ergonomics guidance on individualization within interactive systems, including recommendations on where individualization might be appropriate or inappropriate and how to apply individualization. It focuses on individualization of the software user interface to support the needs of users as individuals or as members of a defined group. It does not recommend specific implementations of individualization mechanisms. It provides guidance on how the various aspects of individualization are made usable and accessible but does not specify which individualizations are to be included within a system.

ISO 9241-210: 2019 "*Ergonomics of human-system interaction - Part 210: Human-centred design for interactive systems*" provides requirements and recommendations for human-centred design principles and activities throughout the life cycle of computer-based interactive systems. It is intended to be used by those managing design processes. It is concerned with ways in which both hardware and software components of interactive systems can enhance human-system interaction. Computer-based interactive systems vary in scale and complexity. Examples include off-the-shelf (shrink-wrap) software products, custom office systems, process control systems, automated banking systems, Web sites and applications, and consumer products such as vending machines, mobile phones, and digital television. Throughout this document, such systems are generally referred to as products, systems, or services, although, for simplicity, sometimes only one term is used.

ISO/TR 9241-810: 2020 "*Ergonomics of human-system interaction - Part 810: Robots, intelligent and autonomous systems*" addresses physically embodied robots, intelligent and autonomous (RIA) systems, such as robots and autonomous vehicles with which users will physically interact; systems embedded within the physical environment with which users do not consciously interact, but which collect data and/or modify the environment within which people live or work such as smart building and, mood-detection; intelligent software tools and agents with which users actively interact through some form of user interface; intelligent software agents which act without active user input to modify or tailor the systems to the user's behavior, task or some other purpose, including providing context specific content/information, tailoring adverts to a user based on information about them, user interfaces that adapt to the cognitive or physiological state; the effect on users resulting from the combined interaction of several RIA systems such as conflicting behaviors between the RIA systems under the same circumstances; the complex system-of-systems and sociotechnical impacts of the use of RIA systems, particularly on society and government. This document identifies where and why ethical issues need to be considered for a wide range of systems and contexts, and as such, it provides information relevant to the broader debate regarding RIA systems.

## 5.1.2 ISO/IEC JTC1/SC42 Artificial intelligence subcommittee

ISO/IEC JTC1 SC42 is a subcommittee under the Joint Technical Committee 1 (JTC1) of ISO and IEC. JTC1 is responsible for the development of international standards in the field of information technology, and SC42 focuses specifically on AI. SC42's primary goal is to develop international standards and guidelines related to AI technologies, addressing both the technical aspects and the societal impact of AI. The committee works on establishing comprehensive standards to ensure that AI is developed in a responsible, transparent, and ethical manner, and that AI systems can be integrated safely and effectively into various industries. ISO/IEC JTC1 SC42 has been instrumental in developing a variety of AI standards; the primary international standards relevant to HAII design are briefly outlined as follows.

ISO/IEC TR 24028:2020 "*Information technology — Artificial intelligence — Overview of Trustworthiness in Artificial Intelligence*" surveys topics related to trustworthiness in AI systems, including approaches to establish trust in AI systems through transparency, explainability, controllability, etc., engineering pitfalls and typical associated threats and risks to AI systems, along with possible mitigation techniques and methods; and approaches to assess and achieve availability, resiliency, reliability, accuracy, safety, security and privacy of AI systems.

ISO/IEC TR 24027:2021 "*Information technology — Artificial intelligence — Bias in AI systems and AI aided decision making*" addresses bias in relation to AI systems, especially about AI-aided decision-making. Measurement techniques and methods for assessing bias are described, with the aim to address and treat bias-related vulnerabilities. All AI system lifecycle phases are in scope, including but not limited to data collection, training, continual learning, design, testing, evaluation, and use.

ISO/IEC 22989:2022 "*Information technology — Artificial intelligence — Artificial intelligence concepts and terminology*" establishes terminology for AI and describes concepts in the field of AI. This document can be used in the development of other standards and in support of communications among diverse, interested parties or stakeholders. This document is applicable to all types of organizations (e.g., commercial enterprises, government agencies, and not-for-profit organizations).

ISO/IEC TR 24368:2022 "*Information technology — Artificial intelligence — Overview of ethical and societal concerns*" provides a high-level overview of AI's ethical and societal concerns. In addition, this document provides information in relation to principles, processes, and methods in this area; it is intended for technologists, regulators, interest groups, and society at large; it is not intended to advocate for any specific set of values (value systems).

ISO/IEC TS 12791:2024 "*Information technology — Artificial intelligence — Treatment of unwanted bias in classification and regression machine learning tasks*" describes how to address unwanted bias in AI systems that use machine learning to conduct classification and regression tasks. This document provides mitigation techniques that can be applied throughout the AI system life cycle in order to treat unwanted bias. This document is applicable to organizations of all types and sizes.

ISO/IEC 5339:2024 "*Information technology — Artificial intelligence — Guidance for AI applications*" provides guidance for identifying the context, opportunities, and processes for developing and applying AI applications. The guidance provides a macro-level view of the AI application context, the stakeholders and their roles, their relationship to the life cycle of the system, and common AI application characteristics and considerations.

ISO/IEC TS 8200:2024 "*Information technology — Artificial intelligence — Controllability of automated artificial intelligence systems*" specifies a basic framework with principles, characteristics, and approaches for the realization and enhancement of automated AI systems' controllability. The following areas are covered: state observability and state transition; control transfer process and cost; reaction to uncertainty during control transfer; verification and validation approaches. This document is applicable to all types of organizations (e.g., commercial enterprises, government agencies, not-for-profit organizations) developing and using AI systems during their whole life cycle.

## 5.2 European Standards

The CEN/CENELEC Guide 8 "*Guidelines for the Development of Harmonized Standards for AI*" is a pivotal framework established by CEN and CENELEC to guide the development of harmonized standards for AI within the European Union (EU). This guide plays a critical role in aligning AI technologies with European values, ethical principles, and legal requirements, particularly in preparation for the EU AI Act.

The CEN/CENELEC Guide 8 focuses on four key objectives. First, it supports the harmonization of AI standards across the EU, fostering interoperability and reducing fragmentation. Second, it promotes the development of ethical AI by embedding principles such as transparency, fairness, and accountability into standardization processes. Third, the guide aims to facilitate innovation and market growth by providing clear guidelines for compliance, reducing barriers for businesses, and ensuring the safety of AI systems. Finally, it addresses the cross-sectoral needs of AI applications, creating standards applicable to industries ranging from healthcare to manufacturing and autonomous transportation.

The CEN/CENELEC Guide 8 introduces several critical features. It aligns standards with EU laws and ethical policies, ensuring that AI systems prioritize user safety and privacy. The guide emphasizes a human-centered approach, advocating for systems that respect user rights, enhance human capabilities, and operate transparently. It adopts a lifecycle perspective, covering all phases of AI development, from data collection and model training to deployment and decommissioning. Additionally, the guide stresses inclusivity and accessibility, ensuring equitable technology access for diverse user groups, and promoting interoperability, allowing AI systems to integrate seamlessly across sectors and platforms.

The guide encourages collaboration among stakeholders, including standardization bodies, industry leaders, policymakers, and researchers, to create robust and inclusive standards. It integrates testing, validation, and certification frameworks to ensure compliance while advocating for continuous improvement through iterative updates. The impact of the guide is profound. Embedding ethical principles into AI standards promotes trustworthy AI aligned with European values. It facilitates legal and regulatory compliance with the EU AI Act, reduces uncertainty for businesses, and lowers barriers to entry for small and medium-sized enterprises (SMEs). Furthermore, the guide positions Europe as a global leader in ethical AI standardization, influencing international policy discussions.

## 5.3 National standards

### 5.3.1 US Standards

The National Institute of Standards and Technology (NIST) plays a critical role in shaping the development and adoption of AI standards in the United States. As a federal agency within the U.S. Department of Commerce, NIST is tasked with advancing technology, innovation, and measurement science, which includes creating frameworks and guidelines for AI to ensure its safe, ethical, and effective implementation. The HAII standards are part of the NIST initiatives to establish guidelines for designing and evaluating effective, trustworthy, and user-centered AI systems. While NIST has not released a single comprehensive "human-AI interaction standard," it has published several reports and frameworks addressing key aspects of HAII. Below are the highlights:

NIST Special Publication (SP) 1270 – *Managing Bias in AI*. This document outlines a framework for identifying, understanding, and mitigating bias in AI systems, which directly impacts HAII. It emphasizes transparency, fairness, and accountability in AI design to ensure equitable and trustworthy interactions.

NIST *AI Risk Management Framework* (AI RMF 1.0). Released in January 2023, this framework focuses on building trustworthy AI systems by addressing risks related to safety, security, fairness, and privacy. It highlights human-centric design principles for ensuring effective communication, user experience, and explainability in HAIIs. The framework is intended for use across industries to guide responsible AI adoption.

NIST's efforts contribute significantly to the development of trustworthy and human-centric AI systems, providing actionable guidelines for industries, policymakers, and developers. These guidelines aim to ensure that AI enhances human capabilities, operates responsibly, and aligns with societal values.

### 5.3.2 British Standards

The BSI is the national standards body of the United Kingdom. Established in 1901, it develops and publishes technical standards for a wide range of industries, including construction, healthcare, manufacturing, information technology, and more. As one of the world's leading standardization organizations, BSI supports innovation, sustainability, and international trade by offering solutions for improved performance and risk management.

BS 8611:2016 "*Guide to the Ethical Design and Application of Robots and Robotic Systems*" is a British Standard developed by BSI. It provides a framework for identifying and addressing ethical hazards associated with robots and robotic systems to ensure their safe and ethical use. The primary goal of BS 8611 is to offer guidance to developers, manufacturers, and users of robots on the ethical risks posed by robotic systems. It emphasizes the need to integrate ethical considerations into the design, development, and deployment stages to minimize potential harm and maximize societal benefits.

BS 8611 identifies various ethical risks related to robots and robotic systems, grouped into categories such as physical hazards, psychological hazards, social hazards, and environmental hazards, and introduces a structured process for identifying and mitigating ethical hazards. The standard provides recommendations for incorporating ethical considerations into the design process, including ensuring transparency in how robots function and make decisions; supporting accountability by clearly defining who is responsible for a robot's actions and outcomes; maintaining respect for privacy, autonomy, and dignity of individuals interacting with robotic systems; addressing bias and fairness in how robots operate and impact society.

BS 8611 is one of the first standards to address the ethical implications of robotics, recognizing the growing societal impact of these technologies. It provides organizations with a practical guide to responsibly develop and deploy robots while respecting human rights, fostering trust, and ensuring safety.

### 5.3.3 Chinese Standards

China places great importance on the development of HAII standards. In July 2020, the Standardization Administration of China, together with the Central Cyberspace Administration, the National Development and Reform Commission, the Ministry of Science and Technology, and the Ministry of Industry and Information Technology, jointly issued the "*Guidelines for the Construction of the National New Generation Artificial Intelligence Standard System.*" In this document, HAII standards are identified as key technical standards within the AI standard system. These standards prioritize the development of intelligent perception technologies, such as integrated scene perception, eye tracking, and three-dimensional input. They also focus on dynamic recognition technologies, including facial recognition, gesture recognition, and handwriting recognition, as well as multimodal interaction standards, such as voice interaction, emotional interaction, body interaction, brain-computer interaction, and full-duplex interaction.

In China, the National Technical Committee 7 on Ergonomics of SAC (SAC/TC 7) is responsible for developing national standards related to human-system interaction. SAC/TC 7 oversees the development of standards for HAII design, including the *Ergonomics of Human-System Interaction—Guidelines for Designing Human-Intelligence System Interaction.*

## 5.4 IEEE Standards

The IEEE is a leading broad technology industry standard setter. IEEE standards provide a fair, competitive environment for global innovation, protect public safety, health and welfare, and contribute to a more sustainable future. The IEEE Robotics and Automation Society (IEEE RAS) has established a working group that is developing human-robot interaction-related standards (Craig S., 2022). The main IEEE standards applicable to HAII design are briefly introduced as follows.

IEEE P3107 "*Human-Robot Interaction Standard Terminology*" defines terms related to human-robot interaction in service, social, educational, industrial, and research robot applications.

IEEE P3108 "*Recommended Practices for Human Subject Research in Human-Robot Interaction Design*" outlines the best practices and requirements for designing human subject experiments in human-robot interaction research.

IEEE P7000 "*Model Process for Addressing Ethical Concerns During System Design*" provides a framework for addressing ethical issues during the design phase of systems, including AI systems. It emphasizes the need for ethical impact assessments and integrating ethical considerations into the development lifecycle.

IEEE P7001 "*Transparency of Autonomous Systems*" focuses on ensuring transparency in autonomous systems, which is crucial for enabling users to understand how these systems operate. It defines guidelines for creating systems that are interpretable and explainable, reducing the "black box" problem in AI.

IEEE P7002 "*Data Privacy Process*" is concerned with ensuring privacy in AI systems. It provides guidelines for managing data privacy, including the collection, processing, and sharing of data, to ensure that user privacy rights are respected and protected.

IEEE P7003 "*Algorithmic Bias Considerations*" focuses on identifying and mitigating biases in algorithms used in AI systems. It addresses the ethical implications of bias in AI, ensuring that systems do not perpetuate or exacerbate social inequalities, discrimination, or unfair treatment.

IEEE P7004 "*Design and Application of Autonomous Systems to Enhance Safety*" focuses on the safety of autonomous systems, including AI-based systems. It outlines best practices for ensuring the safety of users and the environment when using these systems, particularly in high-risk areas such as healthcare, transportation, and military applications.

IEEE P7005 "*Standard for Autonomous and Intelligent Systems: Job Classification and Workforce Impact*" addresses the social and economic impact of AI on the workforce. It focuses on the ethical implications of automation, including job displacement, retraining opportunities, and how AI systems can be designed to complement human workers rather than replace them.

IEEE P7006 "*Standard for Ethical AI in Healthcare*" focuses on the specific ethical challenges related to the use of AI in healthcare. It provides guidelines for ensuring that AI systems in healthcare respect patient privacy, ensure equity in treatment, and support informed decision-making.

IEEE P7007 "*Ontological Standard for Ethically Driven Robotics and Automation Systems*" explores the ethical principles and ontologies that should guide the development of autonomous systems and robots. It focuses on the moral responsibility of developers and the ethical implications of using robots in society.

IEEE P7008 "*Standard for Ethically Driven Measures for Robots, Intelligent, and Autonomous Systems*" provides guidance for developers and ethicists involved in the design of autonomous intelligent systems, describing typical measures (currently in use or potentially created), including necessary concepts, functions, and benefits, to establish and ensure the adoption of ethically driven methods to design robots, intelligent, and autonomous systems.

IEEE P7010 "*Well-Being Metrics for Autonomous and Intelligent Systems*" is concerned with ensuring that autonomous systems contribute positively to human well-being. It introduces metrics to assess how AI systems can improve the quality of life for users and society at large.

# 6 Corporate Guidelines for the Design of HAII

As AI technology becomes increasingly integrated into products, companies like Apple, Google, and Microsoft have introduced their own design guidelines for interacting with AI systems.

## 6.1 Microsoft's Guidelines for HAII

Microsoft's Guidelines for HAII was published at CHI 2019 in May of 2019. Microsoft Research has formulated 18 design guidelines to assist developers in creating AI systems that interact effectively with humans. These guidelines cover various interaction phases, including initial engagement, ongoing interaction, error handling, and long-term use. They aim to ensure AI systems are intuitive, trustworthy, and responsive to user needs (Amershi et al., 2019).

### 6.1.1 Initial stage

(1) Make clear what the system can do. Help the user understand what the AI system can do.

(2) Make clear how well the system can do what it can do. Help the user understand how often the AI system may make mistakes.

### 6.1.2 During interaction

(3) Time services based on context. Time when to act or interrupt based on the user's current task and environment.

(4) Show contextually relevant information. Display information relevant to the user's current task and environment.

(5) Match relevant social norms. Ensure the experience is delivered in a way that users would expect, given their social and cultural context.

(6) Mitigate social biases. Ensure the AI system's language and behaviors do not reinforce undesirable and unfair stereotypes and biases.

### 6.1.3 When wrong

(7) Support efficient invocation. Make it easy to invoke or request the AI system's services when needed.

(8) Support efficient dismissal. Make it easy to dismiss or ignore undesired AI system services.

(9) Support efficient correction. Make it easy to edit, refine, or recover when the AI system is wrong.

(10) Scope services when in doubt. Engage in disambiguation or gracefully degrade the AI system's services when uncertain about a user's goals.

(11) Make clear why the system did what it did. Enable the user to access an explanation of why the AI system behaved as it did.

### 6.1.4 Over time

(12) Remember recent interactions. Maintain short-term memory and allow the user to make efficient references to that memory.

(13) Learn from user behavior. Personalize the user's experience by learning from their actions over time.

(14) Update and adapt cautiously. Limit disruptive changes when updating and adapting the AI system's behaviors.

(15) Encourage granular feedback. Enable the user to provide feedback indicating their prefer¬ences during regular interaction with the AI system.

(16) Convey the consequences of user actions. Immediately update or convey how user actions will impact future behaviors of the AI system.

(17) Provide global controls. Allow the user to globally customize what the AI system monitors and how it behaves.

(18) Notify users about changes. Inform the user when the AI system adds or updates its capabilities.

## 6.2 Google's People + AI Guidebook

In May 2019, Google released the "People + AI Guidebook". This guidebook is based on data and insights from Google's product teams and academic research. Google organizes its content into distinct concepts that developers must continuously keep in mind. These are: User Needs + Defining Success, Data Collection + Evaluation, Mental Models, Explainability + Trust, Feedback + Control, and Errors + Graceful Failure.

### 6.2.1 User needs + Defining success

(1) Find the intersection of user needs & AI strengths. Solve a real problem in ways in which AI adds unique value.

(2) Assess automation vs. augmentation. Automate tasks that are difficult, unpleasant, or where there's a need for scale; and ideally ones where people who currently do it can agree on the "correct" way to do it. Augment tasks that people enjoy doing, that carry social capital, or where people don't agree on the "correct" way to do it.

(3) Design & evaluate the reward function. The "reward function" is how an AI defines successes and failures. Deliberately design this function with a cross-functional team, optimizing for long-term user benefits by imagining the downstream effects of your product. Share this function with users when possible.

### 6.2.2 Data collection + Evaluation

(4) Plan to gather high-quality data from the start. Data is critical to AI, but more time and resources are often invested in model development than data quality. You'll need to plan ahead as you gather and prepare data to avoid the effects of poor data choices further downstream in the AI development cycle.

(5) Translate user needs into data needs. Determine the type of data needed to train your model. You'll need to consider predictive power, relevance, fairness, privacy, and security.

(6) Source your data responsibly. Whether using pre-labeled data or collecting your own, it's critical to evaluate your data and their collection method to ensure they're appropriate for your project.

(7) Prepare and document your data. Prepare your dataset for AI, and document its contents and the decisions that you made while gathering and processing the data.

(8) Design for labelers & labeling. For supervised learning, having accurate data labels is crucial to getting useful output from your model. Thoughtful design of labeler instructions and UI flows will help yield better quality labels and therefore better output.

(9) Tune your model. Once your model is running, interpret the AI output to ensure it is aligned with product goals and user needs. If it's not, thenexplore potential issues with your data.

### 6.2.3 Mental Models

(10) Set expectations for adaptation. AI allows for more systems to adapt, optimize, and personalize for users, and probability-based user experiences have become more common over time. Building on the familiarity of existing mental models can help users feel comfortable.

(11) Onboard in stages. When introducing users to an AI-powered product, explain what it can do, what it can't do, how it may change, and how to improve it.

(12) Plan for co-learning. People will give feedback to AI products, which will adjust the models and change how people interact with them — which will change the machine learning models further. Users' mental models will similarly change over time.

(13) Account for user expectations of human-like interaction. People are more likely to have unachievable expectations for products that they assume have human-like capabilities. It's important to communicate the algorithmic nature and limits of these products to set realistic user expectations and avoid unintended deception.

### 6.2.4 Explainability + Trust

(14) Help users calibrate their trust. Because AI products are based on statistics and probability, the user shouldn't trust the system completely. Rather, based on system explanations, the user should know when to trust the system's predictions and when to apply their own judgement.

(15) Plan for trust calibration throughout the product experience. Establishing the right level of trust takes time. AI can change and adapt over time, and so will the user's relationship with the product.

(16) Optimize for understanding. In some cases, there may be no explicit, comprehensive explanation for the output of a complex algorithm. Even the developers of the AI may not know precisely how it works. In other cases, the reasoning behind a prediction may be knowable, but difficult to explain to users in terms they will understand.

(17) Manage influence on user decisions. AI systems often generate output that the user needs to act on. If, when, and how the system calculates and shows confidence levels can be critical in informing the user's decision-making and calibrating their trust.

### 6.2.5 Feedback + Control

(18) Align feedback with model improvement. Clarify the differences between implicit and explicit feedback, and ask useful questions at the right level of detail.

(19) Communicate value & time to impact. Understand why people give feedback so you can set expectations for how and when it will improve their user experience.

(20) Balance control & automation. Give users control over certain aspects of the experience and allow them to easily opt out of giving feedback.

### 6.2.6  Errors + Graceful failure

(21) Define "errors" & "failure". What the user considers an error is deeply connected to their expectations of the AI system. For example, a recommendation system that's useful 60% of the time could be seen as a failure or a success, depending on the user and the purpose of the system. How these interactions are handled establishes or corrects mental models and calibrates user trust.

(22) Identify error sources. With AI systems, errors can come from many places, be harder to identify and appear to the user and to system creators in non-intuitive ways.

(23) Provide paths forward from failure. AI capabilities can change over time. Creating paths for users to take action in response to the errors they encounter encourages patience with the system, keeps the user-AI relationship going, and supports a better overall experience.

## 6.3  Apple's Human Interface Guidelines for Machine Learning

Apple also released its Human Interface Guidelines for Machine Learning in June 2019 at WWDC'19. The guidelines focus on the application of Apple's design principles in AI products incorporating machine learning. The guidelines are broken up into two main themes: the inputs of a system and the outputs of a system. For inputs, the guidelines focus on Explicit Feedback, Implicit Feedback, Calibration, and Corrections. For outputs, the guidelines cover Mistakes, Multiple Options, Confidence, Attribution, and Limitations.

### 6.3.1  Explicit feedback

(1) Request explicit feedback only when necessary. People must take action to provide explicit feedback, so it's best to avoid requesting it if possible.

(2) Always make providing explicit feedback a voluntary task. You want to communicate that explicit feedback can help improve the experience without making people feel that providing it is mandatory.[

(3) Don't ask for both positive and negative feedback. Good suggestions don't require any feedback, so you don't want to imply that people need to give positive feedback on all the results they like. Instead, give people the opportunity to provide negative feedback on results they don't like so that you can improve the experience.

(4) Use simple, direct language to describe each explicit feedback option and its consequences. Avoid using imprecise terms such as dislike because such terms don't convey consequences and can be hard to translate.

(5) Add icons to an option description if it helps people understand it. Icons can help clarify or emphasize part of an option description. Avoid using an icon by itself, because it might not be clear enough to communicate granularity or consequences.

(6) Consider offering multiple options when requesting explicit feedback. Providing multiple options can give people a sense of control and help them identify unwanted suggestions and remove them from your app.

(7) Act immediately when you receive explicit feedback and persist the resulting changes. When you react immediately to feedback and show that your app remembers it, you build people's trust in the value of providing it.

(8) Consider using explicit feedback to help improve when and where you show results. Explicit feedback on when and where to show results can help you fine-tune the experience people have in your app.

## 6.3.2 Implicit feedback

(9) Always secure people's information. Implicit feedback can gather potentially sensitive user information, so you must be particularly careful to maintain strict controls on user privacy.

(10) Help people control their information. It's important to tell people how your app gets and shares their information and to give people ways to restrict the flow of their information.

(11) Don't let implicit feedback decrease people's opportunities to explore. Implicit feedback tends to reinforce people's behavior, which can improve the user experience in the short term but may worsen it in the long term.

(12) When possible, use multiple feedback signals to improve suggestions and mitigate mistakes. Implicit feedback is indirect, so it can be difficult to discern a person's actual intent in the information you gather.

(13) Consider withholding private or sensitive suggestions. If your app receives implicit feedback related to private or sensitive topics, avoid offering recommendations based on that feedback.

(14) Prioritize recent feedback. People's tastes can change frequently, so base your recommendations on recent implicit feedback.

(15) Use feedback to update predictions on a cadence that matches the person's mental model of the feature.

(16) Be prepared for changes in implicit feedback when you make changes to your app's UI. Even small UI changes can lead to noticeable changes in the amount and types of implicit feedback.

(17) Beware of confirmation bias. Implicit feedback is constrained by what people can actually see and do in your app and other apps — it rarely gives your insight into new things they might like to do. Avoid relying solely on implicit feedback to inform your results.

## 6.3.3 Calibration

(18) Always secure people's information. During calibration, people may provide sensitive information, and you must make sure it remains secure.

(19) Be clear about why you need people's information. Typically, calibration is required before people can use a feature, so it's essential that they understand the value of providing their information.

(20) Collect only the most essential information. Designing a unique experience that requests a minimal amount of information can make people more comfortable participating in the process and increase their trust in your app.

(21) Avoid asking people to participate in calibration more than once. Also, it's best when calibration occurs early in the user experience. As people continue using your app or feature, you can use implicit or explicit feedback to evolve your information about them without asking them to participate again.

(22) Make calibration quick and easy. An ideal calibration experience makes it easy for people to respond, without compromising the quality of the information they provide.

(23) Make sure people know how to perform calibration successfully. After people decide to participate in calibration, give them an explicit goal and show their progress towards it.

(24) Immediately provide assistance if progress stalls. In this situation, it's crucial to give people actionable recommendations that quickly get them back on track.

(25) Confirm success. The moment people successfully complete calibration, reward their time and effort by giving them a clear path towards using the feature.

(26) Let people cancel calibration at any time. Make sure you give people an easy way to cancel the experience at any point and avoid implying any judgment about their choice.

(27) Give people a way to update or remove information they provided during calibration. Letting people edit their information gives them more control and can lead them to have greater trust in your app.

## 6.3.4 Corrections

(28) Give people familiar, easy ways to make corrections. When your app makes a mistake, you don't want people to be confused about how to correct it. You can avoid causing confusion by showing the steps your app takes as it performs the automated task.

(29) Provide immediate value when people make a correction. Reward people's effort by instantly displaying corrected content, especially when the feature is critical or you're responding to direct user input. Also, be sure to keep the updates consistent so people don't have to make the same corrections again.

(30) Let people correct their corrections. Sometimes, people make a correction without realizing that they've made a mistake. As you do with all corrections, respond immediately to an updated correction and persist the update.

(31) Always balance the benefits of a feature with the effort required to make a correction. People may not mind when a feature that automates a task makes a mistake, but they're likely to stop using the feature if it seems easier to perform the task themselves.

(32) Never rely on corrections to make up for low-quality results. Although corrections can reduce the impact of a mistake, depending on them may erode people's trust in your app and reduce the value of your feature.

(33) Learn from corrections when it makes sense. A correction is a type of implicit feedback that can give you valuable information about ways your app doesn't meet people's expectations.

(34) When possible, use guided corrections instead of freeform corrections. Guided corrections suggest specific alternatives, so they require less user effort; freeform corrections don't suggest specific alternatives, so they require more input from people.

## 6.3.5  Mistakes

(35) Understand the significance of a mistake's consequences. Show empathy by providing corrective actions or tools that match the seriousness of the mistake.

(36) Make it easy for people to correct frequent or predictable mistakes. If you don't give people an easy way to correct mistakes, they can lose trust in your app.

(37) Continuously update your feature to reflect people's evolving interests and preferences and help avoid mistakes. Ideally, people don't have to do any work to benefit from improvements in your app.

(38) When possible, address mistakes without complicating the UI. Balance a pattern's effect on the UI with its potential for compounding the mistake.

(39) Be especially careful to avoid mistakes in proactive features. A proactive feature — like a suggestion based on people's behaviors — promises valuable results without asking people to do anything to get them. However, because people don't request a proactive feature, they often have less patience with its mistakes.

(40) As you work on reducing mistakes in one area, always consider the effect your work has on other areas and overall accuracy. Use what you know about people's preferences to help you determine the areas to work on.

## 6.3.6  Multiple Options

(41) Prefer diverse options. When possible, balance the accuracy of a response with the diversity of multiple options. Providing different types of options helps people choose the one that they prefer and can also suggest new items that might interest them.

(42) In general, avoid providing too many options. People must evaluate each option before making a choice, so more options increase cognitive load.

(43) List the most likely option first. When you know how your confidence values correlate with the quality of your results, you might use them to rank the options. If it makes sense in your app, select the first option by default so people can quickly achieve their goals without reading through every option.

(44) Make options easy to distinguish and choose. When options look similar, help people distinguish between them by providing a brief description of each one and taking particular care to highlight the differences.

(45) Learn from selections when it makes sense. When it doesn't adversely affect the user experience, use the feedback to refine the options you provide and increase the chance that you'll present the most likely option first.

### 6.3.7 Confidence

(46) Know what your confidence values mean before you decide how to present them. People may forgive low-quality results from critical or complementary features but presenting low-quality results in a prominent way is likely to erode trust in your app.

(47) In general, translate confidence values into concepts that people already understand. Simply displaying a confidence value doesn't necessarily help people understand how it relates to a result.

(48) In situations where attributions aren't helpful, consider ranking or ordering the results in a way that implies confidence levels. If you must display confidence directly, consider expressing it in terms of semantic categories.

(49) In scenarios where people expect statistical or numerical information, display confidence values that help them interpret the results.

(50) Whenever possible, help people make decisions by conveying confidence in terms of actionable suggestions. Understanding people's goals is key to expressing confidence in ways that help them make decisions.

(51) Consider changing how you present results based on different confidence thresholds. If high or low levels of confidence have a meaningful impact on the ways people can experience the results, it's a good idea to adapt your presentation accordingly.

(52) When you know that confidence values correspond to result quality, you generally want to avoid showing results when confidence is low. Especially when a feature is proactive and can make unbidden suggestions, poor results can cause people to be annoyed and even lose trust in the feature.

### 6.3.8 Attribution

(53) Consider using attributions to help people distinguish among results.

(54) Avoid being too specific or too general. Overly specific attributions can make people feel like they have to do additional work to interpret the results, whereas overly general attributions typically don't provide useful information.

(55) Keep attributions factual and based on objective analysis. To be useful, an attribution needs to help people reason about a result; you don't want to provoke an emotional response.

(56) In general, avoid technical or statistical jargon. In most situations, using percentages, statistics, and other technical jargon doesn't help people assess the results you provide.

### 6.3.9 Limitations

(57) Help people establish realistic expectations. When a limitation may have a serious effect on user experience but happens rarely, consider making people aware of the limitation before they use your app or feature.

(58) Demonstrate how to get the best results. If you don't provide guidance for using a feature, people may assume it'll do everything they want.

(59) Explain how limitations can cause unsatisfactory results. Ideally, your feature can recognize and describe the reasons for poor results to make people aware of the limitations and help them to adjust their expectations.

(60) Consider telling people when limitations are resolved. When you update your app to remove a limitation, you might want to notify people so that they can adjust their mental model of your feature and return to interactions they'd previously avoided.

# 7 Available Tools for Implementing HAII Standards and Guidelines

The proliferation of AI applications in consumer and enterprise contexts has created a pressing need for standardized HAII standards and guidelines. These standards and guidelines emphasize principles such as transparency, interpretability, fairness, and user control. Implementing these standards and guidelines, however, requires tools that address the technical and user experience challenges inherent in AI system design. The available tools and their effectiveness in implementing these interaction standards and guidelines are outlined below.

## 7.1 +Explainability Tools

Explainability is one of the most critical elements in HAII, as it enables users to understand the rationale behind AI decisions. This understanding foster trust, improves decision-making processes, and ensures accountability. Explainability tools are designed to elucidate the behavior of AI models, particularly those classified as "black-box" systems, such as deep learning and ensemble models. The following section provides an overview of the key explainability tools

### 7.1.1 LIME (Local Interpretable Model-Agnostic Explanations)

LIME is one of the most widely used explainability tools for machine learning. It works by approximating the predictions of a complex, black-box model using simpler interpretable models (Ribeiro, Singh, & Guestrin, 2016).

Mechanism: LIME perturbs the input data (e.g., modifies text or images) and observes the changes in the model's predictions. Based on these perturbations, LIME fits a local interpretable surrogate model, such as a linear regression or decision tree, that mimics the behavior of the black-box model for a specific instance.

Applications: In healthcare, LIME can explain the predictions of diagnostic models by highlighting which features (e.g., specific symptoms or test results) influenced the decision most. In financial systems, it can explain credit approval or denial decisions by identifying the most significant input factors.

Strengths: One notable advantage is its model-agnostic nature, allowing it to work seamlessly with any machine learning algorithm. Additionally, it is highly user-friendly, offering clear and intuitive visualizations of feature contributions that are easily interpretable by humans.

Limitations: One key limitation is its instability, as the surrogate model can produce inconsistent results when executed multiple times. Additionally, it has a local scope, offering explanations only for individual instances rather than providing insights into the overall behavior of the model.

### 7.1.2 Google's What-If Tool

The What-If Tool, integrated with TensorBoard, is an interactive tool for testing and debugging machine learning models. It allows users to analyze model performance and fairness interactively without writing code (Google AI.,2019).

Mechanism: Users can modify feature values and observe how these changes affect the model's predictions. The tool also supports counterfactual analysis, which identifies minimal changes needed to achieve a different prediction.

Applications: In fairness analysis, detects biases in models by testing how predictions vary across demographic groups. In debugging, identifies instances where the model performs poorly and allows users to experiment with different inputs.

Strengths: A major advantage is that it requires no coding, making it highly accessible to non-technical users. Furthermore, it offers visual and intuitive features, including interactive visualizations of decision boundaries and prediction changes, enhancing user understanding and engagement.

Limitations: One significant limitation is its limited scalability, as it is best suited for small datasets and simpler models. Additionally, it has a dependency on TensorFlow, as it integrates seamlessly only with TensorFlow-based models.

### 7.1.3 Alibi Explain

Alibi Explain is an open-source Python library specifically designed for explainability in machine learning. It offers multiple techniques, including counterfactual explanations, anchors, and feature importance scoring (Alibi Explain, 2024).

Mechanism: Alibi employs two key approaches to enhance explainability. Counterfactual Explanations generate "what-if" scenarios, demonstrating how small changes to input features could influence the model's predictions. Anchors provide "if-then" rules that clearly outline the reasoning behind a model's decision for a specific input instance.

Applications: In image classification, explains why certain visual elements lead to specific classifications. In text analysis, highlights keywords or phrases that drive sentiment analysis predictions.

Strengths: One of Alibi's key strengths is its versatility, as it supports multiple explanation techniques, making it adaptable to various use cases. Additionally, it is open-source, allowing free access for both academic and commercial purposes, which enhances its accessibility and usability across different domains.

Limitations: However, Alibi does have some notable limitations. It requires a solid understanding of machine learning, which can present a steep learning curve for users without technical expertise. Furthermore, compared to tools like the What-If Tool, Alibi Explain lacks user-friendly, interactive interfaces, which may hinder its user experience for non-technical audiences.

## 7.2 Fairness Tools

Ensuring fairness in AI systems is critical to minimizing biases and preventing discriminatory outcomes. Fairness tools have been developed to identify, quantify, and mitigate biases in datasets and machine learning models, helping organizations align their AI systems with ethical and regulatory standards. This section reviews the most widely used fairness tools, their mechanisms, applications, strengths, and limitations.

## 7.2.1 IBM AI Fairness 360 (AIF360)

IBM AI Fairness 360 is an open-source toolkit designed to evaluate and mitigate bias in machine learning datasets and models. It includes a comprehensive suite of metrics and algorithms for assessing and improving fairness (Bellamy et al., 2018).

Mechanism: AIF360 operates through a three-stage process. The first stage, Bias Detection, provides fairness metrics such as disparate impact, statistical parity difference, and equal opportunity difference to identify potential biases. In the second stage, Bias Mitigation, it employs pre-processing, in-processing, and post-processing techniques to address and reduce bias. Finally, the Evaluation stage compares model performance before and after mitigation to assess the effectiveness of the applied techniques.

Applications: AIF360 is widely used in various fields to address fairness concerns. In healthcare, it helps evaluate biases in diagnostic AI systems across different demographic groups, ensuring equitable outcomes. In finance, it is used to promote fairness in credit scoring algorithms by identifying and mitigating biases related to income levels or ethnicity.

Strengths: AIF360 offers a comprehensive solution by covering a wide range of fairness metrics and mitigation methods, making it suitable for diverse use cases. Additionally, it is highly extensible, with seamless integration support for popular machine learning libraries such as Scikit-learn, enhancing its user experience across different workflows.

Limitations: AIF360 has some notable limitations. It presents a steep learning curve, as effective implementation requires significant technical expertise. Additionally, its functionality is primarily focused on static datasets, offering limited support for dynamic, real-time applications, which can restrict its user experience in certain scenarios.

## 7.2.2 Fairlearn

Fairlearn is an open-source Python toolkit focused on assessing and improving fairness in AI models. It emphasizes ease of integration into existing workflows and interpretability for developers (Bird et al., 2020).

Mechanism: Fairlearn offers two primary functionalities to address fairness in AI systems. First, it computes fairness metrics, such as demographic parity and equalized odds, to evaluate the fairness of model outcomes across different groups. Second, it provides mitigation strategies, including "reduction" algorithms, which aim to simultaneously optimize both fairness and accuracy during model training.

Applications: Fairlearn is particularly useful in fields where fairness is a critical concern. In hiring systems, it helps reduce biases in recruitment algorithms, ensuring equitable treatment of different demographic groups. In education, it is used to analyze and address fairness issues in student performance prediction models, focusing on disparities across genders or ethnicities.

Strengths: One of Fairlearn's key strengths is its developer-friendly design, as it integrates seamlessly with Python-based machine learning workflows, making it easy to adopt. Additionally, it provides interactive visualizations through intuitive dashboards, enabling users to explore and understand fairness metrics effectively.

Limitations: Despite its advantages, Fairlearn has certain limitations. It primarily focuses on in-processing methods, offering fewer tools for pre-processing or post-processing bias mitigation. Furthermore, it faces scalability challenges, making it less suitable for large-scale models or datasets, which can limit its application in more complex scenarios.

### 7.2.3 Themis-ML

Themis-ML is a fairness toolkit specifically designed for use in machine learning pipelines. It provides tools for bias detection and fairness-aware model training (Binns, 2018).

Mechanism: Themis-ML operates by identifying and addressing biases within datasets and models. It measures biases using metrics such as disparate impact and conditional demographic disparity to assess fairness levels. Additionally, it trains models with fairness-aware algorithms, ensuring that outcomes are balanced across sensitive groups.

Applications: Themis-ML is particularly applicable in domains where fairness is critical. In banking, it helps evaluate and mitigate biases in loan approval systems, promoting equitable access to financial resources. In the criminal justice system, it is used to reduce bias in recidivism prediction algorithms, ensuring fairer treatment across different demographic groups.

Strengths: A major strength of Themis-ML is its seamless integration with popular machine learning libraries like Scikit-learn, making it easier to incorporate into existing workflows. Furthermore, its focus on fairness-aware training algorithms ensures that fairness considerations are integrated directly into the model training process.

Limitations: Despite its benefits, Themis-ML has some limitations. The lack of comprehensive documentation can hinder user experience, especially for new users. Additionally, it has a narrow scope, focusing primarily on specific fairness metrics and algorithms, which may limit its applicability to broader fairness concerns or more complex systems.

### 7.2.4 Microsoft Fairness Dashboard

Microsoft's Fairness Dashboard provides a visualization and assessment toolkit for identifying biases in machine learning models. It integrates fairness analysis directly into the model evaluation process (Mishra et al., 2022).

Mechanism: Microsoft's Fairness Dashboard operates by calculating group fairness metrics across sensitive attributes such as gender or race. These metrics help identify potential biases in machine learning models. Additionally, it provides actionable insights to guide bias mitigation during model evaluation and deployment, ensuring fairer outcomes.

Applications: The Fairness Dashboard is particularly useful in applications where fairness plays a critical role. In e-commerce, it ensures that product recommendation algorithms treat users equitably across different demographic groups. In social media, it helps reduce biases in content moderation systems, promoting fairness in automated decision-making processes.

Strengths: A standout strength of the Fairness Dashboard is its visualization capabilities, offering clear and intuitive presentations of fairness metrics that make them accessible to a wide range of stakeholders. Additionally, its cloud integration enables seamless use with Azure Machine Learning services, making it a convenient option for organizations already leveraging Microsoft's ecosystem.

Limitations: Despite its utility, the Fairness Dashboard has some notable limitations. It focuses primarily on fairness assessment, with limited tools for actively mitigating bias. Furthermore, it is platform-dependent, as it is specifically designed for use within Microsoft Azure environments, which may restrict its adoption outside that ecosystem.

## 7.3  Transparency Tools

Transparency is a foundational principle in HAII, enabling users to understand how AI systems operate, the rationale behind their decisions, and their limitations. Transparency tools facilitate the documentation, visualization, and communication of the inner workings of AI systems, making them accessible to stakeholders ranging from developers to end-users. This section outlines key tools for promoting transparency in AI, focusing on their mechanisms, applications, strengths, and limitations.

### 7.3.1  DARPA's XAI (Explainable AI) Framework

The Defense Advanced Research Projects Agency (DARPA) initiated the Explainable AI (XAI) program to address the challenges of understanding and trusting complex machine learning systems. The XAI framework emphasizes creating AI systems that are explainable to humans while maintaining high-performance levels (Gunning & Aha, 2019).

The DARPA XAI framework is centered around three primary goals to enhance the interpretability of AI systems. It aims to develop understandable models that possess inherent interpretability, such as decision trees or rule-based systems, making their decision-making processes transparent. Additionally, it emphasizes the creation of explanation interfaces that utilize post-hoc techniques to provide insights into black-box models like deep learning systems. Lastly, the framework prioritizes human-centric explanations that are concise, actionable, and tailored to meet the cognitive needs and expertise of users, ensuring the explanations are practical and easy to understand.

Mechanism: The DARPA XAI framework employs a range of techniques to improve the transparency of complex AI systems. Methods such as saliency maps, attention mechanisms, and surrogate models are used to make opaque systems more interpretable. Furthermore, the framework incorporates user feedback to refine and enhance the quality of the explanations, ensuring they align with the specific requirements of end-users and stakeholders. This iterative approach allows the framework to provide explanations that are both meaningful and practical.

Applications: The DARPA XAI framework has found applications in critical domains where interpretability and transparency are essential. In the military, it supports transparent decision-making for autonomous systems, ensuring that decisions made by AI can be trusted and understood by human operators. In healthcare, the framework is used to explain AI-driven recommendations for diagnostics and treatment plans, enabling medical professionals to assess and validate the AI's decisions, thereby improving trust and reliability in sensitive scenarios.

Strengths: One of the major strengths of the DARPA XAI framework is its comprehensive approach, as it covers both intrinsic interpretability, where models are designed to be inherently transparent, and post-hoc explainability, which focuses on making black-box models interpretable after they have been developed. Additionally, the framework places a strong emphasis on user experience, ensuring that the explanations provided are tailored to the needs of human users, making them actionable and understandable in real-world applications.

Limitations: Despite its strengths, the DARPA XAI framework has some notable limitations. Its methods often involve significant complexity, requiring advanced technical expertise to implement effectively, which may limit its accessibility to non-expert users. Moreover, its explanation techniques face scalability challenges, as they may struggle to efficiently handle very large models or datasets, potentially limiting their applicability in large-scale AI systems or high-dimensional data environments.

### 7.3.2 Google Explainable AI

Google Explainable AI offers a suite of tools designed to enhance the transparency of machine learning models. It focuses on feature attribution and visualization techniques to make AI models interpretable (Google AI.,2019).

Mechanism: Google Explainable AI employs various techniques to enhance model transparency and interpretability. These include methods like Integrated Gradients, saliency maps, and feature importance scores, which identify and visualize the inputs that most influence a model's predictions. Additionally, the What-If Tool allows users to perform counterfactual analysis by altering input variables and observing how these changes affect the model's predictions, providing a deeper understanding of model behavior.

Applications: This toolset has diverse applications across industries. In e-commerce, it is used to explain why specific products are recommended to users, ensuring transparency in personalized recommendations. In content moderation, it helps stakeholders understand how AI-driven filtering decisions are made, enabling better assessment and refinement of moderation algorithms.

Strengths: One of the key strengths of Google Explainable AI is its interactive visualizations, which allow users to explore and analyze model behavior in an intuitive and hands-on manner. Moreover, it integrates seamlessly with TensorFlow and other Google Cloud models, providing a streamlined workflow for developers working within Google's ecosystem.

Limitations: Despite its advantages, the toolset has certain limitations. It is largely platform-dependent, being primarily designed for use within Google Cloud environments, which may restrict its applicability for users outside this ecosystem. Additionally, the complexity of some of its visualizations requires technical expertise, making it challenging for non-technical users to fully leverage its capabilities.

### 7.3.3 Model Cards for Model Reporting

Model Cards, proposed by Mitchell et al. (2019), are standardized documents that describe a machine learning model's intended use, performance, and limitations. They promote transparency by providing clear, accessible information for stakeholders.

Mechanism: Model Cards are designed to improve transparency in machine learning by following a structured reporting template. They include key sections such as Intended Use Cases, which describe the model's purpose and suitable applications, and a Training Data Summary, which provides details about the dataset's composition and highlights any potential biases. Additionally, they document Performance Metrics, reporting accuracy, fairness, and other relevant measures across different data subgroups. The cards also address Ethical Considerations, outlining risks, limitations, and the potential for misuse to guide responsible deployment.

Applications: Model Cards are widely used across industries to enhance understanding and accountability in AI systems. In finance, they are employed to document credit scoring models, ensuring compliance with regulatory requirements by providing detailed insights into the model's fairness and performance. In healthcare, they are used to summarize AI-driven diagnostic tools, helping stakeholders, including medical professionals and patients, better understand the model's capabilities and limitations.

Strengths: A key strength of Model Cards is their standardization, which ensures a uniform and consistent format for documenting machine learning models. This consistency aids in making model details easier to compare and evaluate across different systems. Additionally, their focus on accessibility enables them to effectively communicate technical information in a user-friendly manner, making them valuable tools for both technical and non-technical stakeholders.

Limitations: Despite their benefits, Model Cards have certain limitations. Creating them requires manual effort, as significant time and expertise are needed to compile accurate and comprehensive documentation. Furthermore, they are inherently static, providing a snapshot of the model at a given time, which limits their ability to deliver real-time insights or offer interactive explanations for ongoing system behavior.

# 8 Case Studies of Human-Centered AI Interaction Design

Human-centered AI emphasizes the creation of AI systems that prioritize human well-being and empowerment. At the core of this approach are HAII design standards, which provide guidelines for developing AI systems that are not only effective but also ethical and trustworthy. This section analyzes case studies from diverse fields, including healthcare, autonomous vehicles, and customer service, to demonstrate how these design standards are applied and developed in real-world AI systems. These examples highlight the practical implementation of principles such as usability, transparency, and ethical alignment, showcasing the impact of thoughtful design on both user experience and societal outcomes.

## 8.1 Guidelines for Designing Patient-Centered AI Tools

Artificial Intelligence (AI) has the potential to revolutionize healthcare by providing personalized, efficient, and scalable solutions to enhance patient care. However, for AI tools to be effective and beneficial, particularly in the medical context, they must be designed with a patient-centered approach. Patient-centered design ensures that the technology meets the specific needs, preferences, and values of the patients it is intended to serve. IBM Watson for Oncology exemplifies how AI can assist oncologists by analyzing patient data and providing evidence-based treatment recommendations (IBM Watson., 2024). The guidelines for designing such tools are outlined below.

### 8.1.1 Usability and Human-Centered Design

Patient-centered AI tools must prioritize usability to facilitate their adoption by clinicians and their meaningful use by patients. Watson for Oncology tailors treatment options based on a patient's medical history, diagnosis, and preferences, ensuring that recommendations align with individual needs. Interfaces should be designed to align with existing clinical workflows, minimizing disruption. These tools must present actionable insights in a clear and concise manner, ensuring that both clinicians and patients can easily interpret and act on AI-generated recommendations. For instance, an AI tool predicting cardiovascular risk should provide recommendations in a user-friendly format, highlighting actionable interventions such as lifestyle changes or treatment adjustments.

### 8.1.2 Transparency and Explainability

Transparency is critical for fostering trust in AI systems, especially in high-stakes healthcare settings. The clinicians and patients must understand the reasoning behind AI predictions to make informed decisions. Tools should provide clear explanations for their outputs, including the key factors driving predictions. For example, a readmission risk predictor should specify which patient variables, such as recent hospitalizations or lab results, contributed most significantly to the risk score. Moreover, communicating the limitations and confidence levels of predictions helps prevent over-reliance on AI systems.

### 8.1.3 Fairness and Bias Mitigation

Bias in AI systems can perpetuate inequities in healthcare, disproportionately affecting underserved populations. Biases in training datasets—such as underrepresentation of certain demographic groups—can lead to inaccurate predictions for these populations. To address this, patient-centered AI tools must be trained on diverse datasets that reflect the populations they are intended to serve. Regular fairness audits should be conducted to detect and mitigate biases during development and deployment. For example, an AI tool designed to detect skin cancer must include training data representing a variety of skin tones to ensure accuracy across demographic groups.

### 8.1.4 Safety and Robustness

Ensuring the safety and robustness of AI systems is paramount in clinical medicine. Deploying models trained in one context to another without proper validation can result in significant errors. AI tools should undergo rigorous testing across diverse clinical settings to ensure consistent performance. Additionally, systems should include safeguards that alert users when predictions fall outside the model's validated range. For example, a medication dosage predictor should flag unusual dosage recommendations for further review by clinicians, reducing the risk of errors.

### 8.1.5 Seamless Integration into Clinical Workflows

Effective integration into clinical workflows is essential for the adoption of AI tools. Poorly integrated systems can lead to frustration among clinicians and hinder their use. Patient-centered tools should be designed to complement existing workflows by embedding AI insights directly into electronic health records (EHRs) and presenting them at the point of care. For instance, an early sepsis detection tool should deliver real-time alerts and actionable insights directly within the EHR interface, ensuring accessibility and reducing workflow disruptions.

### 8.1.6 Continuous Learning and Adaptation

AI tools must be designed to adapt to new data and medical advancements. Watson for Oncology continuously updates its database with the latest clinical research to ensure its recommendations remain relevant and evidence-based. Similarly, patient-centered AI tools should incorporate mechanisms for continuous learning, enabling them to improve accuracy, relevance, and usability over time. Regular updates and validation are necessary to maintain their reliability and clinical utility.

## 8.2 Guidelines for Designing Human-Driver Interactions

The rise of automated vehicles has transformed the dynamics of human-driver interactions, introducing complexities in managing control transitions, situational awareness, and ethical decision-making. The relationship between human drivers and automated systems must be carefully designed to ensure safety, usability, and accountability (Abadie, A., & Desmarais, F., 2019). This section explores the challenges associated with these interactions and outlines guidelines to address them, emphasizing the role of machine ethics and human-centered design in the development of automated vehicles.

### 8.2.1 Enhancing Situation Awareness

One of the primary challenges in human-driver interaction design is ensuring that drivers remain sufficiently aware of their surroundings, even when the automated system is in control. Prolonged reliance on automation can lead to "out-of-the-loop" behavior, where drivers lose situation awareness and are unable to respond effectively when manual intervention is required. To address the issue of situation awareness, automated systems should provide drivers with continuous updates about the vehicle's actions and the surrounding environment. For example, visual or auditory feedback can be used to keep the driver informed about lane changes, speed adjustments, or nearby hazards. Situation awareness can be improved by incorporating predictive models that alert drivers to potential risks before they occur.

### 8.2.2 Promoting Appropriate Trust

Trust is a critical factor in the adoption of automated vehicles. Drivers may either over-rely on automation, leading to complacency, or distrust the system, resulting in underutilization. Building appropriate levels of trust requires systems to be both transparent and reliable. Automated vehicles should provide clear explanations for their decisions, especially in unusual scenarios. For instance, a system could display messages such as "Reducing speed due to detected obstacle" to help the driver understand its actions. There is a need for systems to communicate their limitations, such as weather conditions under which automation may be less effective.

### 8.2.3 Integrating Ethical Decision Frameworks

Automated vehicles face ethical dilemmas in unavoidable harm situations, requiring decisions about prioritizing the safety of passengers versus pedestrians. Ethical decision-making should be explicitly incorporated into the design of automated vehicle systems. Decision-making frameworks should prioritize harm minimization and align with societal norms. Moreover, the system's ethical considerations should be made transparent to the driver, fostering accountability and public trust.

### 8.2.4 Designing Effective Control Transitions

Smooth transitions between automated and manual control are essential for safety but can be difficult to achieve. Drivers may not have enough time to regain full control of the vehicle, particularly in high-speed or complex situations. Clear, multi-modal alerts (e.g., visual, auditory, and haptic) are essential for notifying drivers when manual intervention is required. Additionally, systems should provide a countdown or buffer period during which drivers can prepare to take control. For example, the system might display a message such as "Manual control required in 10 seconds" along with vibration feedback to ensure the driver is fully engaged.

## 8.3 Best Practices for Human-Agent Interactions in Service Systems

As artificial intelligence becomes more prevalent in service systems, organizations face the challenge of designing interactions that meet customer expectations and align with ethical standards. AI systems in customer service, such as chatbots and virtual agents, are expected to provide accurate, timely, and personalized responses while maintaining transparency and fairness (Adam & Wessel, 2020). The best practices for human-agent interactions in service systems, focusing on improving user experience and fostering trust in AI-driven services are outlined.

### 8.3.1 Transparency in AI Communication

Transparency is a cornerstone of effective AI-customer interactions. Customers shall be informed when they are interacting with an AI system rather than a human agent. Clear communication about the AI's capabilities, limitations, and purpose helps manage user expectations and builds trust. For example, virtual agents should explicitly state their role at the beginning of an interaction, such as: "I am a virtual assistant designed to help you with basic inquiries." Transparency also involves providing explanations for AI decisions. For instance, if an AI recommends a specific product or solution, it should explain the factors influencing that recommendation. Such practices ensure that users feel informed and maintain trust in the service system.

### 8.3.2 Personalization and Context Awareness

Personalization is key to delivering high-quality service experiences. AI systems should understand and adapt to individual customer preferences and past interactions. AI systems should leverage customer data responsibly to provide context-aware responses tailored to the user's needs. For example, an AI system in e-commerce can recommend products based on a customer's browsing history and preferences. However, personalization must be balanced with privacy considerations. Customers should have control over the data used by AI systems, and service providers must ensure compliance with data protection regulations.

### 8.3.3 Ethical Considerations in AI Design

Ethical considerations are critical in the design of AI service systems. AI systems must be trained on diverse datasets to avoid discriminatory outcomes and ensure that all customer segments receive equitable treatment. In addition, ethical AI design involves safeguarding customer data and providing transparency about how data is collected, stored, and used. Organizations should establish clear guidelines for data use and ensure that customers can opt out of data sharing if desired. These practices not only enhance trust but also align with broader societal expectations of ethical AI deployment.

### 8.3.4 Seamless Escalation to Human Agents

While AI systems can handle routine queries effectively, there are situations where human involvement is necessary. The service systems should allow for seamless escalation from AI to human agents. Customers should be able to request human assistance easily, particularly when the issue is complex or requires empathy and nuanced judgment. For instance, chatbots should include an option to transfer the conversation to a human agent when the AI system detects frustration or dissatisfaction. This ensures that customers receive the level of service they need while maintaining a positive overall experience.

### 8.3.5 Continuous Learning and Improvement

AI service systems must be designed with mechanisms for continuous learning and improvement. The customer interactions provide valuable feedback that can be used to enhance the system's performance. AI systems should be able to learn from past interactions to improve their accuracy and relevance over time. Additionally, organizations should regularly evaluate their AI systems to identify potential biases, errors, or inefficiencies. Human oversight is essential in this process to ensure that the system evolves in a way that aligns with customer needs and organizational goals.

# 9 Challenges and Future Directions in Human-AI Interaction Design Standards

AI systems are increasingly embedded in critical domains such as healthcare, transportation, education, and customer service, where HAII plays a pivotal role. The quality of these interactions impacts trust, usability, and overall effectiveness. However, the development of standards for HAII design has lagged behind the rapid pace of AI innovation, resulting in fragmented efforts and inconsistent practices across industries. This chapter identifies the major challenges in establishing these standards and outlines future directions to overcome these barriers, ultimately supporting the design of trustworthy, user-centered AI systems.

## 9.1 Challenges in Human-AI Interaction Design Standards

### 9.1.1 Lack of Universal Standards

One of the primary challenges in HAII design is the absence of globally accepted standards. While organizations such as the ISO and IEEE have worked to create guidelines for AI development, there is no universal standard that is widely adopted across industries. This issue is compounded by regional differences in cultural norms, regulations, and ethical values, which lead to fragmented efforts. For instance, the definition of "trustworthiness" in AI varies from country to country, as different cultures and legal systems prioritize different aspects of trust and accountability (González et al., 2020). The lack of a unified global standard makes it difficult to establish common principles for designing AI systems that can be universally understood and trusted.

### 9.1.2 Complexity of Human-AI Interactions

AI systems are integrated into a wide array of contexts, each with its own requirements and constraints. For example, the human-AI interaction in autonomous vehicles differs drastically from that in healthcare applications, where machine learning models assist in diagnosing diseases. This variety of applications, combined with the dynamic nature of AI systems—particularly machine learning systems that adapt based on new data—creates a major challenge in standardizing human-AI interactions. A universal framework for designing these interactions that can accommodate all possible use cases remains elusive. As AI systems become more autonomous and complex, it is increasingly difficult to create standards that can keep up with the evolving nature of these technologies (Dastin, 2018).

### 9.1.3 Bias and Fairness Issues

Bias and fairness are central issues when designing AI systems that interact with humans. AI systems are often trained using historical data, which may reflect existing societal biases, such as racial, gender, or socioeconomic prejudices. These biases, if not addressed, can lead to discriminatory outcomes. For example, facial recognition software has been shown to have higher error rates for women and people of color, perpetuating biases in its decisions (Angwin et al., 2016). Addressing bias requires not only technical solutions such as bias correction algorithms but also a shared understanding of fairness, which can differ across cultural and demographic boundaries. The challenge lies in developing standards that can ensure AI systems are fair and equitable for all users, regardless of their background (Eubanks, 2018).

### 9.1.4 Transparency vs. Usability

Transparency is crucial in building trust in AI systems. Users need to understand how decisions are made by the AI to feel confident in using these technologies. However, there is a tradeoff between providing detailed explanations of AI decisions and maintaining the system's usability. Excessive transparency—such as providing overly technical details about how algorithms work—can overwhelm users and reduce the overall user experience. The challenge is to strike the right balance between transparency and simplicity, ensuring that users can understand the key factors influencing AI decisions without being bogged down by unnecessary complexity (Miller, 2019). This is particularly important in critical domains like healthcare, where users must trust AI decisions but may lack the technical expertise to comprehend every aspect of an AI model.

### 9.1.5 Trust and Accountability

Trust is fundamental to human-AI interaction. Users must trust AI systems for them to be accepted and effectively utilized. However, defining trustworthiness in AI is a difficult task, as it involves subjective measures such as reliability, predictability, and fairness. Moreover, accountability for AI decisions remains unclear in many instances. In cases where AI systems make errors or cause harm, it is often difficult to assign responsibility. The lack of clear accountability mechanisms poses significant challenges to establishing standards that users can rely on. Trust and accountability standards must be carefully defined and integrated into AI system design (Langer & Aizenman, 2021).

### 9.1.6 Ethical Dilemmas

AI technologies often raise ethical concerns, such as the potential erosion of user autonomy, privacy, and control over personal data. In healthcare, for example, AI systems may prioritize cost-saving measures over patient outcomes, leading to ethical dilemmas. These ethical considerations complicate the creation of universal standards, as different cultures and legal frameworks may have conflicting views on issues like privacy, consent, and autonomy. Developing AI standards that adequately address these ethical challenges is essential for creating systems that respect human rights and values (Binns, 2018). However, integrating ethics into AI design is a complex task that requires collaboration between AI researchers, ethicists, and policymakers.

### 9.1.7 Rapidly Evolving Technology

The pace at which AI technologies are evolving poses a significant challenge to the development of HAII standards. AI technologies, especially those based on machine learning, are continuously improving, and new breakthroughs are frequently introduced. As a result, standards often lag behind technological advancements, leading to outdated practices and guidelines. This rapid evolution creates a situation where, by the time a standard is established, it may already be irrelevant or inadequate. Keeping standards up-to-date with the latest developments in AI technology is a critical challenge (Brynjolfsson & McAfee, 2014).

### 9.1.8 Cross-Disciplinary Collaboration

Creating effective standards for human-AI interaction requires input from various fields, including AI development, human-computer interaction (HCI), psychology, ethics, and law. However, collaboration between these disciplines is not always straightforward. The differences in language, methodology, and objectives across fields can create barriers to effective collaboration. AI developers may not always have the expertise to understand the psychological or ethical implications of their designs, while ethicists may lack technical knowledge about how AI systems work. Overcoming these interdisciplinary challenges is essential for creating comprehensive and effective HAII standards (Shneiderman, 2020).

## 9.2 Future Directions in Human-AI Interaction Design Standards

### 9.2.1 Development of Global Frameworks

International collaboration is essential for the creation of unified standards. Organizations such as ISO, and IEEE should work together to develop frameworks that accommodate regional differences while maintaining a common foundation.

### 9.2.2 Dynamic and Adaptive Standards

Future standards must be flexible and adaptable to keep pace with technological advancements. Incorporating feedback loops and iterative updates can help ensure relevance over time. Real-world deployment data and user feedback should be leveraged to refine standards continuously.

### 9.2.3 Context-Specific Standards

Instead of universal guidelines, standards should be tailored to specific applications. For instance, autonomous vehicle standards should prioritize safety and reliability, while healthcare standards might emphasize privacy and accuracy.

### 9.2.4 Bias Auditing and Mitigation Frameworks

Comprehensive methods for detecting and mitigating bias in AI systems are critical. Standards should require AI systems to document and disclose the demographic distribution of training data and their performance metrics across different user groups.

### 9.2.5 Explainable and Interpretable AI

Standards for explainability should balance technical transparency with user comprehension. Tiered explanations—offering varying levels of detail based on user expertise—can enhance both usability and trust.

### 9.2.6 Ethical and Inclusive AI Standards

Ethics should be embedded in the design process, ensuring AI systems respect user autonomy, privacy, and societal values. For example, standards should require AI systems to provide options for users to opt out of or override automated decisions.

### 9.2.7 Human-Centered AI Metrics

Developing measurable metrics for evaluating HAIIs is essential. Metrics such as trust, usability, fairness, and psychological safety can guide the development of effective standards. For example, "calibrated trust" metrics can measure whether users appropriately trust an AI system based on its capabilities.

### 9.2.8 Regulatory and Legal Alignment

Design standards should align with emerging AI regulations, such as the EU's AI Act or the U.S. AI Bill of Rights. Certification processes for compliance with these standards can enhance accountability and enforcement.

### 9.2.9 Proactive Security and Privacy Standards

Robust security and privacy guidelines should be integrated into HAII standards to protect sensitive user data. For example, standards could mandate end-to-end encryption for AI systems handling personal or financial information.

# 10 Conclusion

The rapid development of AI has transformed the nature of human-computer interactions, necessitating the establishment of robust design standards to ensure effective, ethical, and human-centered AI solutions. By fostering interdisciplinary collaboration, integrating ethical considerations, and aligning with regulatory frameworks, HAII design standards can support the creation of human-centered AI systems that not only meet user needs but also contribute positively to society.

As artificial intelligence (AI) systems become increasingly integrated into daily life, the need for effective and ethical HAII design grows exponentially. Designing AI systems that are intuitive, trustworthy, and user-centered requires more than technical expertise—it demands a deep understanding of human behavior, cognitive processes, and usability principles. To meet these challenges, continuous collaboration between AI developers and human factors experts is essential.

Collaboration plays a crucial role in bridging the gap between advanced technology and practical usability. While AI developers excel in areas such as machine learning algorithms, data processing, and system implementation, their technical expertise often lacks a user-centric perspective. This can result in systems that are highly functional but challenging for users to understand or operate. Human factors experts bring valuable insights into user behavior, cognitive processes, and interaction design principles, ensuring that AI systems are intuitive and aligned with human needs and capabilities. By working collaboratively, AI developers and human factors specialists can simplify complex AI functionalities, proactively identify and address potential user challenges, and enhance the overall user experience through thoughtful and intuitive design. This synergy ensures that AI systems are not only technically robust but also accessible, efficient, and user-friendly.

Collaboration in addressing ethical and trust Issues. Ethical concerns, such as algorithmic bias, privacy, and transparency, are critical in HAII. AI developers are often focused on optimizing system performance, while human factors experts emphasize trust, inclusivity, and fairness. Collaboration is vital to mitigate biases in AI systems through diverse datasets and user-centered evaluation, design interfaces that are transparent and explainable, helping users trust AI decisions, address privacy concerns by incorporating user consent and data control mechanisms into design standards.

Collaboration is essential in addressing ethical and trust-related challenges in HAII. Critical issues such as algorithmic bias, privacy, and transparency require a balanced approach that combines technical expertise with a focus on human values. While AI developers often prioritize optimizing system performance, human factors experts emphasize the importance of trust, inclusivity, and fairness. By working together, these disciplines can mitigate biases in AI systems using diverse datasets and user-centered evaluation methods. Collaborative efforts can also ensure the design of transparent and explainable interfaces, enabling users to understand and trust AI decisions. Furthermore, privacy concerns can be effectively addressed by incorporating user consent and data control mechanisms into design standards. This interdisciplinary approach not only enhances the ethical alignment of AI systems but also fosters user confidence and acceptance, creating a foundation for responsible and trustworthy AI.

Collaboration is crucial in establishing universal design standards for HAII, as it requires input from both technical and human-centric disciplines. Human factors experts play a key role in ensuring these standards prioritize usability, accessibility, and inclusivity, addressing the diverse needs of users. Meanwhile, AI developers contribute their expertise to ensure that the guidelines are technically feasible, scalable, and aligned with the capabilities of AI systems. By working together, these disciplines can create comprehensive and practical standards that balance human-centered design principles with technical requirements, paving the way for more effective, inclusive, and user-friendly AI solutions.

The future of HAII depends on seamless collaboration between AI developers and human factors experts. By combining technical expertise with insights into human behavior and usability, this partnership can ensure that AI systems are not only powerful and innovative but also accessible, ethical, and user-friendly. To build AI systems that truly enhance human capabilities and foster trust, we must prioritize ongoing collaboration across these disciplines, ensuring that HAII evolves to meet the needs of society effectively.

Author's Introduction

Dr. Chaoyi Zhao, Research Professor, Laboratory of Human Factors and Ergonomics, China National Institute of Standardization. Research field: Ergonomics, human-computer interaction, AI interaction, standardization. E-mail: zhaochy@cnis.ac.cn.